\documentstyle[amssymb,floats,aps,prb,epsf]{revtex}
\epsfclipon
\begin{document}
\twocolumn[\hsize\textwidth\columnwidth\hsize\csname
@twocolumnfalse\endcsname

\draft
\title{Unusual charge transport and spin response of doped bilayer
triangular antiferromagnets}

\author{Ying Liang and Tianxing Ma}
\address{Department of Physics, Beijing Normal University,
Beijing 100875, China}

\author{Shiping Feng}
\address{Department of Physics, Beijing Normal University, Beijing
100875, China\\
Interdisciplinary Center of Theoretical Studies, Chinese Academy
of Sciences, Beijing 100080, China\\
National Laboratory of Superconductivity, Chinese Academy of
Sciences, Beijing 100080, China}
\date{Received December 16, 2002}
\maketitle
\begin{abstract}
Within the $t$-$J$ model, the charge transport and spin response
of the doped bilayer triangular antiferromagnet are studied by
considering the bilayer interaction. Although the bilayer
interaction leads to the band splitting in the electronic
structure, the qualitative behaviors of the physical properties
are the same as in the single layer case. The conductivity spectrum
shows the low-energy peak and unusual midinfrared band, the
temperature dependent resistivity is characterized by the
nonlinearity metallic-like behavior in the higher temperature
range, and the deviation from the metallic-like behavior in the
lower temperature range, and the commensurate neutron scattering
peak near the half-filling is split into six incommensurate peaks
in the underdoped regime, with the incommensurability increases
with the hole concentration at lower dopings, and saturates at
higher dopings.
\end{abstract}
]


\section{Introduction}

It has become clear in the past ten years that the most remarkable
expression of the nonconventional physics of doped cuprates is
found in the normal-state \cite{n1,n2}. These unusual normal-state
properties are due to the special microscopic conditions: (1) Cu
ions situated in a square-planar arrangement and bridged by oxygen
ions (CuO$_{2}$ plane), (2) weak coupling between neighboring
layers, and (3) doping in such a way that the Fermi level lies near
the middle of the Cu-O $\sigma^{*}$ bond, where {\it one common
feature} is the {\it square-planar} Cu arrangement \cite{n1,n2}. In
the underdoped and optimally doped regimes, it has been shown from
the transport experiments \cite{n1,n3} that the ratio of the c-axis
and in-plane resistivity $R=\rho_{c}(T)/\rho_{ab}(T)$ ranges from
$R\sim 100$ to $R>10^{5}$, this reflects that the charged carriers
are tightly confined to the CuO$_{2}$ planes. This large magnitude
of the resistivity anisotropy also leads to the general notion that
the physics of these materials is almost entirely two-dimensional
(2D), and can be well described by a single CuO$_{2}$ plane
\cite{n4}. However, this physical picture seems to be incompatible
with the fact that the superconducting transition temperature
T$_{c}$ is closely related to the number of CuO$_{2}$ planes per
unit cell, with single layer materials of a family generically
having lower T$_{c}$ than bilayer or trilayer materials
\cite{n1,n2}. Furthermore, some essential differences of the
magnetic behaviors between doped single layer and bilayer cuprates
have been found \cite{n5,n6}. In the underdoped regime, it has been
shown from the experiments that only incommensurate neutron
scattering peaks for the single layer lanthanum cuprate are
observed \cite{n5}, however, both low-energy incommensurate neutron
scattering peaks and high-energy commensurate [$\pi$,$\pi$]
resonance for the bilayer yttrium cuprate in the normal state are
detected \cite{n6}. These experimental results highlight the
importance of some sort of coupling between the CuO$_{2}$ planes
within a unit cell. On the other hand, the bilayer band splitting
in the doped bilayer cuprates was shown by the band calculation
\cite{n7}, and clearly observed \cite{n8} recently by the
angle-resolved-photoemission spectroscopy in the underdoped and
overdoped bilayer cuprates. This bilayer band splitting is due to
a nonvanishing intracell coupling. Moreover, the magnitude of the
bilayer splitting is constant over a large range of dopings
\cite{n9}.

However, many materials with the Cu arrangements on 2D non-square
lattices have been synthetized \cite{n10,n11,n12}. In particular,
it has been found from the experiments that there is a class of
doped cuprates, RCuO$_{2+\delta}$, R being a rare-earth element,
where the Cu ions are not arranged on a square-planar, but on a
{\it triangular-planar lattice} \cite{n10}, therefore allowing a
test of the geometry effect on the normal-state properties, while
retaining some other unique special microscopic features of the
Cu-O bond. Since the strong electron correlation is common for
both doped square and triangular antiferromagnets, it is expected
that the unconventional normal-state properties existing in the
doped square antiferromagnet may also be seen in the doped
triangular antiferromagnet. However, within the $t$-$J$ model, we
\cite{n13,n14} have discussed the charge transport and spin
response of the doped single layer triangular antiferromagnet in
the underdoped regime, and found that the normal-state properties
of the doped single layer triangular lattice antiferromagnet are
much different from these of the doped single layer square lattice
antiferromagnet since the strong geometry effect in the triangular
lattice system. The conductivity spectrum shows the unusual
behavior at low energies and anomalous midinfrared band separated
by the charge-tranfer gap, in contrast with the doped square
lattice antiferromagnet, the resistivity exhibits a nonlinearity in
temperatures, and the commensurate neutron scattering peak near the
half-filling is split into six incommensurate scattering peaks away
from the half-filling. Considering these highly unusual charge
transport and spin response in the underdoped regime, a natural
question is what is the effect of the intracell coupling on the
charge transport and spin response of the doped bilayer triangular
antiferromagnet. On the other hand, the undoped bilayer triangular
antiferromagnet is a good candidate for a 2D quantum system with
the resonating valence bond spin liquid due to the strong geometry
spin frustration and quantum interference effect between the layers
\cite{n15}. This spin liquid state would be particularly attractive,
given the intensive work on the spin liquid states on the square
lattice, especially in connection with the superconductivity of
doped cuprates \cite{n4}. Moreover, it has been shown that the
doped and undoped triangular antiferromagnets present a pairing
instability in an unconventional channel \cite{n16}. In this paper,
we apply the fermion-spin theory \cite{n17} to study the charge
transport and spin response of the doped bilayer triangular
antiferromagnet. We hope that the information from the present work
may induce further experimental works in doped antiferromagnets on
the non-square lattice.

The paper is organized as follows. The theoretical framework is
presented in Sec. II, where the single-particle holon and spinon
Green's functions are calculated based on the $t$-$J$ model by
considering the bilayer interaction. Within this theoretical
framework, we discuss the charge transport of the doped bilayer
triangular antiferromagnet in Sec. III. It is shown that the
conductivity spectrum shows a low-energy peak and unusual
midinfrared band, while the temperature dependent resistivity is
characterized by the nonlinearity metallic-like behavior in the
higher temperature range, and the deviation from the metallic-like
behavior in the lower temperature range. In Sec. IV, the spin
response of the doped bilayer triangular antiferromagnet is studied.
Our result shows that the commensurate neutron scattering peak near
the half-filling is split into six incommensurate peaks in the
underdoped regime, where the incommensurability is doping dependent,
and increases with the hole concentration at lower dopings, and
saturates at higher dopings. Sec. V is devoted to a summary and
discussions. Our results also show that although the bilayer
interaction leads to the band splitting in the electronic structure,
the qualitative behavior of the charge transport and spin response
are the same as in the single layer case \cite{n13,n14}.

\section{Theoretical framework}

As in the doped single layer triangular antiferromagnet
\cite{n4,n16}, the essential physics of the doped bilayer
triangular antiferromagnet is well described by the bilayer $t$-$J$
model on the triangular lattice, and can be expressed as,
\begin{eqnarray}
H&=&-t_{\parallel}\sum_{ai\hat{\eta}\sigma}C_{ai\sigma}^{\dagger}
C_{ai+\hat{\eta}\sigma}-t_{\perp}\sum_{i\sigma}
(C_{1i\sigma}^{\dagger}C_{2i\sigma}+{\rm h.c.})\nonumber \\
&-&\mu\sum_{ai\sigma} C_{ai\sigma }^{\dagger }C_{ai\sigma }
+J_{\parallel}\sum_{ai\hat{\eta}}{\bf S}_{ai}\cdot {\bf
S}_{ai+\hat{\eta}}\nonumber \\
&+&J_{\perp}\sum_i{\bf S}_{1i} \cdot {\bf S}_{2i},
\end{eqnarray}
where the summation within the plane is over all sites $i$, and for
each $i$, over its nearest-neighbor $\hat{\eta}$, $a=1$, $2$ is
plane indices, $C^{\dagger}_{ai\sigma}$ ($C_{ai\sigma}$) is the
electron creation (annihilation) operator,
${\bf S}_{ai}=C_{ai}^{\dagger}{\vec{\sigma}}C_{ai}/2$ are spin
operators with ${\vec{\sigma}}=(\sigma_x,\sigma_y,\sigma_z)$ as
Pauli matrices, and $\mu$ is the chemical potential. The bilayer
$t$-$J$ model (1) is supplemented by the single occupancy local
constraint $\sum_\sigma C_{ai\sigma}^{\dagger}C_{ai\sigma}\leq 1$.
This local constraint reflects the strong electron correlation in
the doped antiferromagnet, and can be treated {\it properly in
analytical form} within the fermion-spin theory \cite{n17} based
on the charge-spin separation,
\begin{eqnarray}
C_{ai\uparrow}=h_{ai}^{\dagger}S_{ai}^{-}, ~~~~
C_{ai\downarrow}=h_{ai}^{\dagger}S_{ai}^{+},
\end{eqnarray}
with the spinless fermion operator $h_{ai}$ keeps track of the
charge (holon), while the pseudospin operator $S_{ai}$ keeps track
of the spin (spinon), and then the low-energy behavior of the
bilayer $t$-$J$ model (1) can be rewritten in the fermion-spin
representation as,
\begin{eqnarray}
H&=&t_{\parallel}\sum_{ai\hat{\eta}}h^{\dagger}_{ai+\hat{\eta}}
h_{ai}(S^{+}_{ai}S^{-}_{ai+\hat{\eta}}+S^{-}_{ai}
S^{+}_{ai+\hat{\eta}})\nonumber \\
&+&t_{\perp}\sum_{i}(h^{\dagger}_{1i}h_{2i}+
h^{\dagger}_{2i}h_{1i})(S^{+}_{1i}S^{-}_{2i}+S^{-}_{1i}S^{+}_{2i})\nonumber \\
&+&\mu\sum_{ai}h^{\dagger}_{ai}h_{ai} +J_{{\parallel}{\rm eff}}
\sum_{ai\hat{\eta}}{\bf S}_{ai}\cdot {\bf S}_{ai+\hat{\eta}}\nonumber \\
&+&J_{\perp{\rm eff}}\sum_i{\bf S}_{1i}\cdot {\bf S}_{2i},
\end{eqnarray}
where $S^{+}_{ai}$ ($S^{-}_{ai}$) is the pseudospin raising
(lowering) operator,
$J_{{\parallel}{\rm eff}}=J_{\parallel}[(1-x)^{2}-
\phi_{\parallel}^{2}]$,
$J_{\perp{\rm eff}}=J_{\perp}[(1-x)^{2}-\phi^{2}_{\perp}]$,
$x$ is the hole doping concentration, and
$\phi_{\parallel}=\langle h_{ai}^{\dagger}h_{ai+\hat{\eta}}\rangle$
and $\phi_{\perp}=\langle h^{\dagger}_{1i}h_{2i}\rangle$ are the
holon particle-hole order parameters. The spinon and holon may be
separated at the mean-field (MF) level, but they are strongly
coupled beyond the MF approximation due to the strong holon-spin
interaction.

In the bilayer system, because there are two coupled planes, then
the energy spectrum has two branches. In this case, the
one-particle holon and spinon Green's functions are matrices, and
are expressed as,
\begin{mathletters}
\begin{eqnarray}
D(i-j,\tau-\tau^{\prime})&=&D_{L}(i-j,\tau-\tau^{\prime})\nonumber \\
&+&\sigma_{x}D_{T}(i-j,\tau-\tau^{\prime}), \\
g(i-j,\tau-\tau^{\prime})&=&g_{L}(i-j,\tau-\tau^{\prime})\nonumber \\
&+&\sigma_{x}g_{T}(i-j,\tau-\tau^{\prime}),
\end{eqnarray}
\end{mathletters}
respectively, where the longitudinal and transverse parts are
defined as,
\begin{mathletters}
\begin{eqnarray}
D_{L}(i-j,\tau-\tau^{\prime})=-\langle T_{\tau}S_{ai}^{+}(\tau)
S_{aj}^{-}(\tau^{\prime})\rangle, \\
g_{L}(i-j,\tau-\tau^{\prime})=-\langle T_{\tau}h_{ai}(\tau)
h_{aj}^{\dagger}(\tau^{\prime})\rangle, \\
D_{T}(i-j,\tau-\tau^{\prime})=-\langle T_{\tau}S_{ai}^{+}(\tau)
S_{a^{\prime}j}^{-}(\tau^{\prime})\rangle, \\
g_{T}(i-j,\tau-\tau^{\prime})=-\langle T_{\tau}h_{ai}(\tau)
h_{a^{\prime}j}^{\dagger}(\tau^{\prime})\rangle,
\end{eqnarray}
\end{mathletters}
with $a\neq a^{\prime}$. At the half-filling, the bilayer $t$-$J$
model (3) is reduced as the bilayer antiferromagnetic Heisenberg
model. It has been shown \cite{n18} that as in the square lattice,
there is indeed the antiferromagnetic long-range-order (AFLRO) in
the ground state of the single layer triangular antiferromagnetic
Heisenberg model, but this AFLRO is destroyed more rapidly with
increasing dopings than on the square lattice due to the strong
geometry frustration. Since the quantum interference effect
between the layers in the bilayer triangular antiferromagnet does
not favor the magnetic order for spins, then AFLRO in the doped
bilayer triangular antiferromagnet is suppressed more fastly than
the doped single layer antiferromagnet, therefore away from the
half-filling, there is no AFLRO for the doped bilayer triangular
antiferromagnet, ${\it i.e.}$, $\langle S_{ai}^{z}\rangle=0$.
Within the fermion-spin formalism, the MF theory of the doped
square antiferromagnet in the underdoped and optimally doped
regimes without AFLRO has been developed \cite{n19}. Following
thier discussions, the MF holon and spinon Green's functions of
the doped bilayer triangular antiferromagnet are obtained as,
\begin{mathletters}
\begin{eqnarray}
g^{(0)}_{L}(k,\omega)&=&{1\over 2}\sum_{\nu=1,2}{1\over \omega-
\xi^{(\nu)}_{k}},\\
g^{(0)}_{T}(k,\omega)&=&{1\over 2}\sum_{\nu=1,2}(-1)^{\nu+1}
{1\over \omega-\xi^{(\nu)}_{k}},\\
D_{L}^{(0)}(k,\omega)&=& {1\over 2}\sum_{\nu=1,2}{B_{k}^{(\nu)}
\over\omega^{2}-(\omega_{k}^{(\nu)})^{2}},\\
D_{T}^{(0)}(k,\omega)&=&{1\over 2}\sum_{\nu=1,2}(-1)^{\nu+1}
{B_{k}^{(\nu)}\over \omega^{2}-(\omega_{k}^{(\nu)})^{2}},
\end{eqnarray}
\end{mathletters}
respectively, where
$B_{k}^{(\nu)}=\lambda[(2\epsilon_{\parallel}\chi^{z}_{\parallel}+
\chi_{\parallel})\gamma_{k}-(\epsilon_{\parallel}\chi_{\parallel}+
2\chi^{z}_{\parallel})]-J_{\perp{\rm eff}}[\chi_{\perp}
+2\chi_{\perp}^z(-1)^{\nu}][\epsilon_{\perp}+(-1)^{\nu}]$,
$\lambda=2ZJ_{{\parallel}{\rm eff}}$, $\epsilon_{\parallel}=1+2
t_{\parallel}\phi_{\parallel}/J_{{\parallel}{\rm eff}}$,
$\epsilon_{\perp}=1+4t_{\perp}\phi_{\perp}/J_{\perp{\rm eff}}$,
$\gamma_{k}=[\cos{k_{x}}+2\cos{(k_{x}/2)}\cos{({\sqrt 3}k_{y}/2)}]
/3$, and $Z$ is the number of the nearest neighbor sites within
the plane, while the MF holon and spinon excitation spectra are
given by,
\begin{mathletters}
\begin{eqnarray}
\xi^{(\nu)}_{k}&=&2Zt_{\parallel}\chi_{\parallel}\gamma_{k}+
2\chi_{\perp}t_{\perp}(-1)^{\nu+1}+\mu, \\
(\omega_{k}^{(\nu)})^{2}&=&A_{1}\gamma_{k}^{2}+A_{2}\gamma_{k}
+A_{3}\nonumber \\
&+&(-1)^{\nu+1}(X_{1}\gamma_{k}+X_{2}),
\end{eqnarray}
\end{mathletters}
with
\begin{mathletters}
\begin{eqnarray}
A_{1}&=&\alpha\epsilon_{\parallel}\lambda^{2}({1\over 2}
\chi_{\parallel}+\epsilon_{\parallel}\chi^{z}_{\parallel}), \\
A_{2}&=&\epsilon_{\parallel}\lambda^{2}[(1-Z)\alpha{1\over Z}
({1\over 2}\epsilon_{\parallel}\chi_{\parallel}+
\chi^{z}_{\parallel})\nonumber \\
&-&\alpha(C^{z}_{\parallel}+{1\over 2} C_{\parallel})
-{1\over 2Z}(1-\alpha)]\nonumber \\
&-&\alpha\lambda J_{\perp {\rm eff}}[\epsilon_{\parallel}
(C_{\perp }^{z}+\chi_{\perp}^{z})+{1\over 2}\epsilon_{\perp}
(C_{\perp}+\epsilon_{\parallel}\chi_{\perp})], \\
A_{3}&=&\lambda^{2}[\alpha (C^{z}_{\parallel}+{1\over 2}
\epsilon^{2}_{\parallel}C_{\parallel})+{1\over 4Z}(1-\alpha)
(1+\epsilon^{2}_{\parallel})\nonumber \\
&-&\alpha\epsilon_{\parallel} {1\over Z}({1\over
2}\chi_{\parallel}+\epsilon_{\parallel}
\chi^{z}_{\parallel})] \nonumber \\
&+&\alpha\lambda J_{\perp {\rm eff}}[\epsilon_{\parallel}
\epsilon_{\perp}C_{\perp}+ 2C_{\perp}^{z}]+{1\over 4}
J_{\perp {\rm eff}}^{2}(\epsilon_{\perp}^{2}+1), \\
X_{1}&=&\alpha\lambda J_{\perp {\rm eff}}[{1\over 2}
(\epsilon_{\perp}\chi_{\parallel}+\epsilon_{\parallel}
\chi_{\perp})+\epsilon_{\parallel}\epsilon_{\perp}
(\chi_{\perp}^{z}+\chi^{z}_{\parallel})], \\
X_{2}&=&-\alpha\lambda J_{\perp {\rm eff}}[{1\over 2}
\epsilon_{\parallel}\epsilon_{\perp}\chi_{\parallel}+
\epsilon_{\perp}(\chi^{z}_{\parallel}+C_{\perp}^{z})+
{1\over 2}\epsilon_{\parallel} C_{\perp}]\nonumber \\
&-&\epsilon_{\perp} J_{\perp {\rm eff}}^{2}/2,
\end{eqnarray}
\end{mathletters}
where the spinon correlation functions
$\chi_{\parallel}=\langle S_{ai}^{+}S_{ai+\hat{\eta}}^{-}\rangle$,
$\chi^{z}_{\parallel}=\langle S_{ai}^{z}S_{ai+\hat{\eta}}^{z}
\rangle$, $\chi_{\perp}=\langle S_{1i}^{+}S_{2i}^{-}\rangle$,
$\chi_{\perp}^{z}=\langle S_{1i}^{z}S_{2i}^{z}\rangle$,
$C_{\parallel}=(1/Z^2)\sum_{\hat{\eta}\hat{\eta^{\prime}}}\langle
S_{ai+\hat{\eta}}^{+}S_{ai+\hat{\eta^{\prime}}}^{-}\rangle$,
$C^{z}_{\parallel}=(1/Z^2)\sum_{\hat{\eta}\hat{\eta ^{\prime }}}
\langle S_{ai+\hat{\eta}}^{z}S_{ai+\hat{\eta^{\prime }}}^{z}
\rangle$, $C_{\perp}=(1/Z)\sum_{\hat{\eta}}\langle S_{2i}^{+}
S_{1i+\hat{\eta}}^{-}\rangle$, and $C_{\perp}^{z}=(1/Z)
\sum_{\hat{\eta}}\langle S_{1i}^{z}S_{2i+\hat{\eta}}^{z}\rangle$.
In order to satisfy the sum rule for the correlation function
$\langle S_{ai}^{+}S_{ai}^{-}\rangle=1/2$ in the absence of AFLRO,
a decoupling parameter $\alpha $ has been introduced in the MF
calculation, which can be regarded as the vertex correction
\cite{n19}. As a result of self-consistent motion of holons and
spinons, all these mean-field order parameters, decoupling
parameter, and the chemical potential are determined
self-consistently.

In this paper we hope to discuss the optical, transport, and
magnetic properties of the doped bilayer triangular
antiferromagnet, then the second-order corrections for holons and
spinons due to the holon-spinon interaction are necessary for the
proper description of the holon motion in the background of the
magnetic fluctuation and spinon motion in the background of the
charged holon fluctuation. Within the fermion-spin theory, the full
holon and spinon Green's functions of the doped single layer
triangular antiferromagnet have been evaluated by considering the
holon-spinon interaction \cite{n13,n14}. According to these
discussions, we obtain the full holon and spinon Green's functions
of the doped bilayer triangular antiferromagnet by the loop
expansion to the second-order as,
\begin{mathletters}
\begin{eqnarray}
g^{-1}(k,\omega)&=&g^{(0)-1}(k,\omega)-\Sigma^{(h)}(k,\omega), \\
D^{-1}(k,\omega)&=&D^{(0)-1}(k,\omega)-\Sigma^{(s)}(k,\omega),
\end{eqnarray}
\end{mathletters}
where $\Sigma^{(h)}(k,\omega)=\Sigma^{(h)}_{L}(k,\omega)+\sigma_{x}
\Sigma_{T}^{(h)}(k,\omega)$ is the second-order holon self-energy
from the spinon pair bubble, with the longitudinal and transverse
parts are obtained as,
\begin{mathletters}
\begin{eqnarray}
\Sigma^{(h)}_{L}(k,\omega)&=&{1\over N^{2}}\sum_{pq}
\sum_{\nu\nu'\nu''}\Xi^{(h)}_{\nu\nu'\nu''}(k,p,q,\omega), \\
\Sigma_{T}^{(h)}(k,\omega)&=&{1\over N^{2}}\sum_{pq}
\sum_{\nu\nu'\nu''}(-1)^{\nu+\nu'+\nu''+1}\nonumber \\
&\times &\Xi^{(h)}_{\nu\nu'\nu''} (k,p,q,\omega),
\end{eqnarray}
\end{mathletters}
respectively, and $\Sigma^{(s)}(k,\omega)=\Sigma^{(s)}_{L}
(k,\omega)+\sigma_{x}\Sigma_{T}^{(s)}(k,\omega)$ is the
second-order spinon self-energy from the holon pair bubble, with the
longitudinal and transverse parts are obtained as,
\begin{mathletters}
\begin{eqnarray}
\Sigma_{L}^{(s)}(k,\omega)&=&{1\over N^2}\sum_{pp^{\prime}}
\sum_{\nu\nu^{\prime}\nu^{\prime\prime}}\Xi^{(s)}_{\nu\nu^{\prime}
\nu^{\prime\prime}}(k,p,p^{\prime},\omega), \\
\Sigma_{T}^{(s)}(k,\omega)&=&{1\over N^2}\sum_{pp^{\prime}}
\sum_{\nu\nu^{\prime}\nu^{\prime\prime}}(-1)^{\nu+\nu^{\prime}+
\nu^{\prime\prime}+1}\nonumber \\
&\times &\Xi^{(s)}_{\nu\nu^{\prime}\nu^{\prime\prime}}
(k,p,p^{\prime},\omega),
\end{eqnarray}
\end{mathletters}
respectively, where
\begin{mathletters}
\begin{eqnarray}
\Xi^{(h)}_{\nu\nu'\nu''}(k,p,q,\omega)&=&{B^{(\nu')}_{q+p}
B^{(\nu)}_{q}\over 32\omega^{(\nu')}_{q+p}\omega^{(\nu)}_{q}}
\left( {Zt_{\parallel}[\gamma_{q+p+k}+\gamma_{q-k}]}\right.\nonumber \\
&+&\left.{t_{\perp}
[(-1)^{\nu+\nu''}+(-1)^{\nu'+\nu''}] }\right)^{2} \nonumber \\
&\times&\left ({F^{(1)}_{\nu\nu'\nu''}(k,p,q)\over\omega+
\omega^{(\nu')}_{q+p}-\omega^{(\nu)}_{q}-\xi^{(\nu'')}_{p+k}}
\right. \nonumber \\
&+&\left.{F^{(2)}_{\nu\nu'\nu''}(k,p,q)\over\omega-
\omega^{(\nu')}_{q+p}+\omega^{(\nu)}_{q}-\xi^{(\nu'')}_{p+k}}
\right. \nonumber \\
&+&\left. {F^{(3)}_{\nu\nu'\nu''}(k,p,q)\over\omega+
\omega^{(\nu')}_{q+p}+\omega^{(\nu)}_{q}-\xi^{(\nu'')}_{p+k}}\right.\nonumber \\
&+&\left.{F^{(4)}_{\nu\nu'\nu''}(k,p,q)\over\omega-\omega^{(\nu')}_{q+p}-
\omega^{(\nu)}_{q}-\xi^{(\nu'')}_{p+k}}\right ), \\
\Xi^{(s)}_{\nu\nu^{\prime}\nu^{\prime\prime}}(k,p, p^{\prime},
\omega) &=&{\frac{B_{k+p}^{(\nu^{\prime\prime})}}
{16\omega_{k+p}^{(\nu^{\prime\prime})}}}
\left(Zt_{\parallel}[\gamma_{p^{\prime}+p+k}+
\gamma_{k-p^{\prime}}]\right.\nonumber \\
&+&\left. t_{\perp}[(-1)^{\nu^{\prime}+\nu^{\prime\prime}}+
(-1)^{\nu+\nu^{\prime\prime}}]\right)^{2}
\nonumber \\
&\times &\left({\frac{L_{\nu\nu^{\prime}\nu^{\prime\prime}}^{(1)}
(k,p,p^{\prime})}{\omega+\xi_{p+p^{\prime}}^{(\nu^{\prime})}-
\xi_{p^{\prime}}^{(\nu)}-\omega_{k+p}^{(\nu^{\prime\prime})}}}\right.\nonumber \\
&-&\left.{\frac{L_{\nu\nu^{\prime}\nu^{\prime\prime}}^{(2)}(k,p,p^{\prime})}
{\omega+\xi_{p+p^{\prime}}^{(\nu^{\prime})}-\xi_{p^{\prime}}^{\nu}+
\omega_{k+p}^{(\nu^{\prime\prime})}}} \right),
\end{eqnarray}
\end{mathletters}
with
\begin{mathletters}
\begin{eqnarray}
F^{(1)}_{\nu\nu'\nu''}(k,p,q)&=&n_{F}(\xi^{(\nu'')}_{p+k})[n_{B}
(\omega^{(\nu)}_{q})-n_{B}(\omega^{(\nu')}_{q+p})]\nonumber \\
&+&n_{B}
(\omega^{(\nu')}_{q+p})[1+n_{B}(\omega^{(\nu)}_{q})], \\
F^{(2)}_{\nu\nu'\nu''}(k,p,q)&=&n_{F}(\xi^{(\nu'')}_{p+k})[n_{B}
(\omega^{(\nu')}_{q+p})-n_{B}(\omega^{(\nu)}_{q})]\nonumber \\
&+&n_{B}
(\omega^{(\nu)}_{q})[1+n_{B}(\omega^{(\nu')}_{q+p})], \\
F^{(3)}_{\nu\nu'\nu''}(k,p,q)&=&n_{F}(\xi^{(\nu'')}_{p+k})[1+n_{B}
(\omega^{(\nu')}_{q+p})+n_{B}(\omega^{(\nu)}_{q})]\nonumber \\
&+&n_{B}
(\omega^{(\nu)}_{q})n_{B}(\omega^{(\nu')}_{q+p}), \\
F^{(4)}_{\nu\nu'\nu''}(k,p,q)&=&[1+n_{B}(\omega^{(\nu)}_{q})]
[1+n_{B}(\omega^{(\nu')}_{q+p})]\nonumber \\
&-&n_{F}(\xi^{(\nu'')}_{p+k})
[1+n_{B}(\omega^{(\nu')}_{q+p})\nonumber \\
&+&n_{B}(\omega^{(\nu)}_{q})], \\
L_{\nu\nu^{\prime}\nu^{\prime\prime}}^{(1)}(k,p,p^{\prime})&=&
n_{F}(\xi_{p+p^{\prime}}^{(\nu^{\prime })})[1-n_{F}
(\xi_{p^{\prime}}^{(\nu)})] -n_{B}(\omega_{k+p}^{(\nu^{\prime
\prime})})\nonumber \\
&\times&[n_{F}(\xi_{p^{\prime}}^{(\nu)})-n_{F}(\xi_{p+
p^{\prime}}^{(\nu^{\prime})})], \\
L_{\nu\nu^{\prime}\nu^{\prime\prime}}^{(2)}(k,p,p^{\prime})&=&
n_{F}(\xi_{p+p^{\prime}}^{(\nu^{\prime})})[1-n_{F}
(\xi_{p^{\prime}}^{(\nu)})]\nonumber \\
&+&[1+n_{B}(\omega_{k+p}^{(\nu^{\prime
\prime})})]\nonumber \\
 &\times&[n_{F}(\xi_{p^{\prime}}^{(\nu)})-
n_{F}(\xi_{p+p^{\prime}}^{(\nu^{\prime})})],
\end{eqnarray}
\end{mathletters}
and $n_{F}(\xi_{k}^{(\nu)})$ and $n_{B}(\omega_{k}^{(\nu)})$ are
the fermion and boson distribution functions, respectively. Within
the fermion-spin theory, the spin fluctuation couples only to
spinons, while the charge fluctuation couples only to holons
\cite{n13,n14}, this is because that the local constraint with no
doubly occupied has been treated properly, and therefore leads to
disappearing of the extra gauge degree of freedom related with the
local constraint under the charge-spin separation. However, the
strong correlation between holons and spinons still is included
self-consistently through the holon's order parameters entering the
spinon's propagator and the spinon's order parameters entering the
holon's propagator, therefore both holons and spinons are
responsible for the optical, transport and magnetic behaviors.

\section{Charge transport}
\begin{figure}[prb]
\epsfxsize=2.5in\centerline{\epsffile{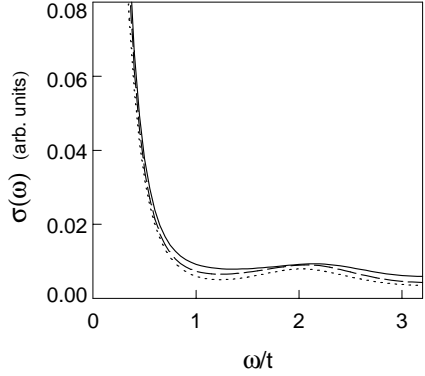}} \caption{The
conductivity at $x=0.12$ (solid line), $x=0.09$ (dashed line), and
$x=0.06$ (dotted line) in $T=0$ for
$t_{\parallel}/J_{\parallel}=2.5$, $t_{\perp
}/t_{\parallel}=0.25$, and $J_{\perp}/J_{\parallel}=0.25$.}
\end{figure}
\begin{figure}[prb]
\epsfxsize=2.5in\centerline{\epsffile{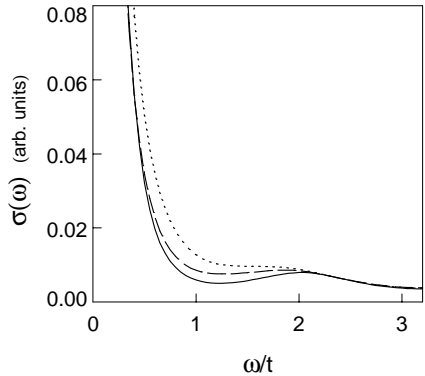}} \caption{The
conductivity in $T=0$ (solid line), $T=0.3J_{\parallel}$ (dashed
line), $T=0.5J_{\parallel}$ (dotted line) at $x=0.06$ for
$t_{\parallel}/J_{\parallel}=2.5$, $t_{\perp}/t_{\parallel}=0.25$,
and $J_{\perp}/J_{\parallel}=0.25$.}
\end{figure}

Among the striking features of the unusual physical properties
stands out the extraordinary charge transport \cite{n1,n2}. The
frequency- and temperature-dependent conductivity is a powerful
probe for systems of interacting electrons, and provides very
detailed informations of the excitations, which interacts with
carriers in the normal-state and might play an important role in
the superconductivity. In the framework of the fermion-spin
theory, the charge fluctuation couples only to holons as mentioned
above, and therefore as in the doped single layer case \cite{n13},
the conductivity of the doped bilayer triangular antiferromegnet
is obtained as,
\begin{eqnarray}
\sigma(\omega)=-{{\rm Im}\Pi^{(h)}(\omega)\over \omega},
\end{eqnarray}
where $\Pi^{(h)}(\omega)$ is the holon current-current correlation
function, and is defined as $\Pi^{(h)}(\tau-\tau')=-\langle T_\tau
j^{(h)}(\tau)j^{(h)}(\tau')\rangle$, with $\tau$ and $\tau'$ are
the imaginary times, and $T_{\tau}$ is the $\tau$ order operator.
Within the Hamiltonian (3), the current density of holons is
obtained by the time derivation of the polarization operator using
Heisenberg's equation of motion as,
\begin{eqnarray}
j^{(h)}&=&2\chi_{\parallel} et_{\parallel}\sum_{ai\hat{\eta}}
\hat{\eta}h_{ai+\hat{\eta}}^{\dagger} h_{ai}\nonumber \\
&+&2\chi_{\perp}e
t_{\perp}\sum_{i}(R_{2i}-R_{1i})(h^{\dagger}_{2i}h_{1i}-
h^{\dagger}_{1i}h_{2i}),
\end{eqnarray}
where $R_{1i}$ ($R_{2i}$) is the lattice site of the plane $1$
(plane $2$), and $e$ is the electronic charge, which is set as the
unit hereafter. This holon current-current correlation function can
be calculated in terms of the full holon Green's function
$g(k,\omega)$. After a straightforward calculation, we obtain
explicitly the conductivity of the doped bilayer triangular
antiferromagnet as,
\begin{eqnarray}
\sigma(\omega)=\sigma^{(L)}(\omega)+\sigma^{(T)}(\omega),
\end{eqnarray}
with the longitudinal and transverse parts are given by,
\begin{mathletters}
\begin{eqnarray}
\sigma^{(L)}(\omega)&=&{1\over N}\sum_{k}[(2Z\chi_{\parallel}
t_{\parallel}\gamma_{sk})^{2}+(2\chi_{\perp} t_{\perp})^{2}]
\nonumber\\
&\times& \int^{\infty}_{-\infty}{d\omega'\over 2\pi}A^{(h)}_{L}
(k,\omega'+\omega)A^{(h)}_{L}(k,\omega')\nonumber \\
&\times&{n_{F}(\omega'+\omega)-
n_{F}(\omega') \over \omega}, \\
\sigma^{(T)}(\omega)&=&{1\over N}\sum_{k}[(2Z\chi_{\parallel}
t_{\parallel}\gamma_{sk})^{2}-(2\chi_{\perp} t_{\perp})^{2}]
\nonumber\\
&\times& \int^{\infty}_{-\infty}{d\omega'\over 2\pi}A^{(h)}_{T}
(k,\omega'+\omega)A^{(h)}_{T}(k,\omega')\nonumber \\
&\times&{n_{F}(\omega'+\omega)- n_{F}(\omega') \over \omega},
\end{eqnarray}
\end{mathletters}
respectively, where
$\gamma_{sk}^{2}=[{\rm sin}(k_{x})+{\rm sin}(k_{x}/2){\rm cos}
(\sqrt{3}k_{y}/2)]^{2}/9+[{\rm cos}(k_{x}/2){\rm sin}
(\sqrt{3}k_{y}/2)]^{2}/3$, $A^{(h)}_{L}(k,\omega)=-2{\rm Im}g_{L}
(k,\omega)$ and $A^{(h)}_{T}(k,\omega)=-2{\rm Im}g_{T}(k,\omega)$
are the holon's longitudinal and transverse spectral functions.
\begin{figure}[prb]
\epsfxsize=2.5in\centerline{\epsffile{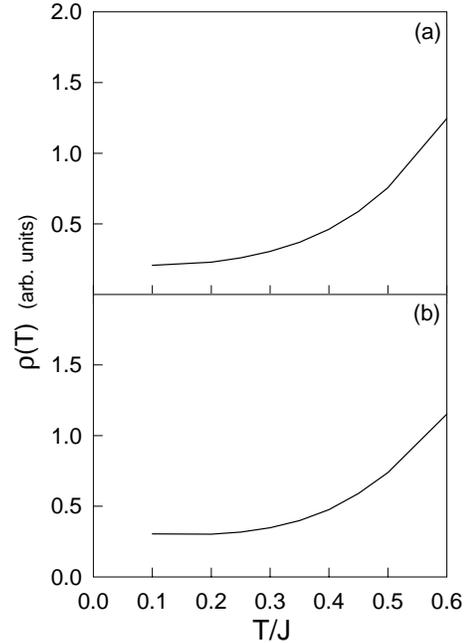}} \caption{The
resistivity at (a) $x=0.12$ and (b) $x=0.06$ for
$t_{\parallel}/J_{\parallel}=2.5$, $t_{\perp
}/t_{\parallel}=0.25$, and $J_{\perp}/J_{\parallel}=0.25$.}
\end{figure}

In Fig. 1, we present the conductivity at doping $x=0.12$ (solid
line), $x=0.09$ (dashed line), and $x=0.06$ (dotted line) for
parameters $t_{\parallel}/J_{\parallel}=2.5$,
$t_{\perp }/t_{\parallel}=0.25$, and $J_{\perp}/J_{\parallel}=0.25$
in temperature $T=0$, where $\sigma(\omega)$ shows a sharp
low-energy peak and the unusual midinfrared band separated by the
charge-transfer gap of the undoped system. This low-energy peak
decays fastly with increasing energy, while the charge-transfer
gap is doping dependent, deceases with increasing dopings, and
vanishes in higher doped regime. For the better understanding of
the optical properties, we also study the conductivity at different
temperatures and the results in $T=0$ (solid line),
$T=0.3J_{\parallel}$ (dashed line), and $T=0.5J_{\parallel}$
(dotted line) at $x=0.06$ for $t_{\parallel}/J_{\parallel}=2.5$,
$t_{\perp }/t_{\parallel}=0.25$, and $J_{\perp}/J_{\parallel}=0.25$
are shown in Fig. 2. We therefore find that the charge-transfer gap
also decreases with increasing temperature, and disappears at
higher temperatures. In the above calculations, we also find that
the conductivity $\sigma(\omega)$ of the doped bilayer triangular
antiferromagnet is essentially determined by its longitudinal part
$\sigma^{(L)}(\omega)$, this is why in the present bilayer system
the conductivity spectrum appears to reflect the single layer
nature of the electronic state \cite{n13}.

Now we turn to discuss the resistivity. The resistivity is closely
related to the conductivity, and can be obtained as
$\rho(T)=1/\lim_{\omega\rightarrow 0}\sigma(\omega)$. We have
performed the calculation for $\rho(T)$, and the results at
$x=0.12$ and $x=0.06$ for $t_{\parallel}/J_{\parallel}=2.5$,
$t_{\perp}/t_{\parallel}=0.25$, and $J_{\perp}/J_{\parallel}=0.25$
are plotted in the Fig. 3(a) and Fig. 3(b), respectively. These
results show that the temperature-dependent resistivity of the
doped bilayer triangular antiferromagnet is characterized by the
nonlinearity metallic-like behavior in the higher temperature
range, and the deviation from the metallic-like behavior in the
lower temperature range. In comparison with the results of the
doped singlar triangular antiferromagnet \cite{n13}, it is shown
that although the bilayer interaction leads to the band splitting
in the electronic structure, the qualitative behavior of the
charge transport is the same as in the single layer case.

Since the $t$-$J$ model is characterized by the competition between
the kinetic energy ($t$) and magnetic energy ($J$). The magnetic
energy $J$ favors the magnetic order for spins, while the kinetic
energy $t$ favors delocalization of holes and tends to destroy the
magnetic order. In the present fermion-spin theory, although both
holons and spinons contribute to the charge transport, the
scattering of holons dominates the charge transport, where the
charged holon scattering rate is obtained from the full holon
Green's function (then the holon self-energy (10) and holon
spectral function) by considering the holon-spinon interaction. In
this case, the unusual behavior in the resistivity is closely
related with this competition. In the underdoped regime, the holon
kinetic energy is much smaller than the magnetic energy in lower
temperatures, therefore the magnetic fluctuation is strong enough
to heavily reduce the charged holon scattering and thus is
responsible for the deviation from the metallic-like behavior in
the resistivity. With increasing temperatures, the holon kinetic
energy is increased, while the spinon magnetic energy is decreased.
In the region where the holon kinetic energy is much larger than
the spinon magnetic energy at high temperatures, the charged holon
scattering would give rise to the metallic-like behavior in the
resistivity.

\section{Spin response}

The antiferromagnetic spin correlation is responsible for the
nuclear magnetic resonance (NMR), nuclear quadrupole resonance
(NQR), and especially for the temperature dependence of the
spin-lattice relaxation rate \cite{n1,n2}. This spin response is
manifested by the dynamical spin structure factor $S(k,\omega)$,
which can be obtained in the present bilayer triangular
antiferromagnet as,
\begin{figure}[prb]
\epsfxsize=2.5in\centerline{\epsffile{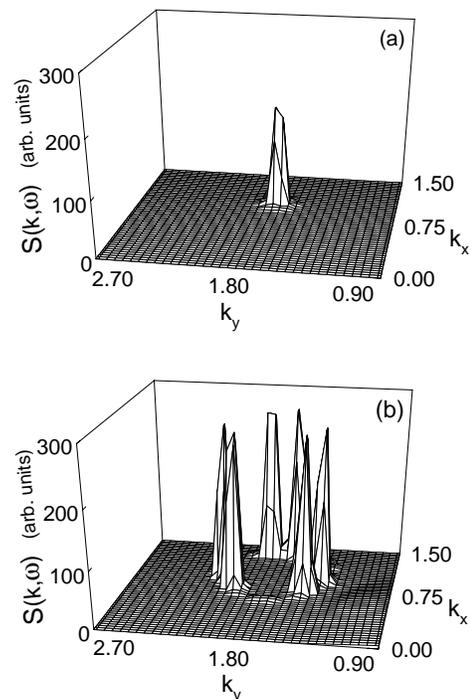}} \caption{The
dynamical spin structure factor in the $(k_{x},k_{y})$ plane at
(a) $x=0.02$ and (b) $x=0.06$ in $T=0.1J_{\parallel}$ and
$\omega=0.05J_{\parallel}$ for $t_{\parallel}/J_{\parallel}=2.5$,
$t_{\perp}/t_{\parallel}=0.2$, and $J_{\perp}/J_{\parallel}=0.2$.}
\end{figure}
\begin{eqnarray}
S(k,\omega)&=&-2[1+n_{B}(\omega)][2{\rm Im}D_{L}(k,\omega)+
2{\rm Im}D_{T}(k,\omega)] \nonumber \\
&=&-{\frac{4[1+n_{B}(\omega)](B_{k}^{(1)})^{2}{\rm Im}
\Sigma_{LT}^{(s)}(k,\omega)}{A+B}},
\end{eqnarray}
where $A=[\omega^{2}-(\omega_{k}^{(1)})^{2}- B_{k}^{(1)}{\rm
Re}\Sigma_{LT}^{(s)}(k,\omega)]^{2}$, $B=[B_{k}^{(1)} {\rm
Im}\Sigma_{LT}^{(s)} (k,\omega)]^{2}$, and ${\rm
Im}\Sigma_{LT}^{(s)}(k,\omega)={\rm Im}\Sigma_{L}^{(s)}
(k,\omega)+{\rm Im}\Sigma_{T}^{(s)}(k,\omega)$ and ${\rm Re}
\Sigma_{LT}^{(s)}(k,\omega)={\rm Re}\Sigma_{L}^{(s)}(k,\omega)+
{\rm Re}\Sigma_{T}^{(s)}(k,\omega)$, while ${\rm Im}
\Sigma_{L}^{(s)}(k,\omega)$ (${\rm Im}\Sigma_{T}^{(s)}(k,\omega)$)
and ${\rm Re}\Sigma_{L}^{(s)}(k,\omega)$ (${\rm
Re}\Sigma_{T}^{(s)} (k,\omega)$) are the imaginary and real parts
of the second order longitudinal (transverse) spinon self-energy
(11), respectively.
\begin{figure}[prb]
\epsfxsize=2.5in\centerline{\epsffile{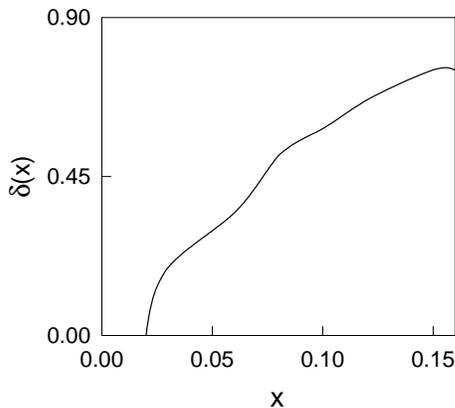}} \caption{The
doping dependence of the incommensurability $\delta(x)$ of the
antiferromagnetic fluctuations.}
\end{figure}

In Fig. 4, we present the result of the dynamical spin structure
factor $S(k,\omega)$ in the ($k_{x},k_{y}$) plane at (a) $x=0.02$
and (b) $x=0.06$ in $T=0.1J_{\parallel}$ and
$\omega=0.05J_{\parallel}$ for $t_{\parallel}/J_{\parallel}=2.5$,
$t_{\perp}/t_{\parallel}=0.2$, and $J_{\perp}/J_{\parallel}=0.2$,
hereafter we use the units of $[2\pi/3,2\pi/3]$. Obviously, there
is a commensurate-incommensurate transition in the spin
fluctuation with dopings. Near the half-filling, the spin
fluctuation is commensurate, and the position of the commensurate
neutron scattering peak is located at the antiferromagnetic wave
vector ${\bf Q}=[1,\sqrt{3}]$. However, this commensurate peak is
split into six peaks in the underdoped regime, while the positions
of these incommensurate peaks are located at
$[(1-\delta_{x}),(\sqrt{3}\pm\delta_{y})]$,
$[(1-\delta'_{x}),(\sqrt{3}\pm \delta'_{y})]$, and
$[(1-\delta''_{x}),(\sqrt{3} \pm\delta''_{y})]$, with
$\sqrt{\delta^{2}_{x}+\delta^{2}_{y}}=\sqrt{(\delta'_{x})^{2}+
(\delta'_{y})^{2}}=\sqrt{(\delta''_{x})^{2}+(\delta''_{y})^{2}}
=\delta$. The calculated dynamical spin structure factor spectrum
has been used to extract the doping dependence of the
incommensurability parameter $\delta(x)$, defined as the deviation
of the peak position from the antiferromagnetic wave vector ${\bf
Q}$, and the result is plotted in Fig. 5, where $\delta(x)$
increases with the hole concentration in lower dopings, but it
saturates at higher dopings, and is qualitatively similar to the
results of the doped single layer triangular antiferromagnet
\cite{n14}. For the further understanding of the magnetic
properties, we have made a series of scans for $S(k,\omega)$ at
different temperatures and energies, and find that as in the
single layer case, the weight of the incommensurate peaks are
broadened and suppressed with increasing temperatures. Moreover,
although the positions of the incommensurate peaks are almost
energy independent, the weight of these peaks decreases with
increasing energy, and tends to vanish at high energies. This
reflects that the inverse lifetime of the spin excitations
increases with increasing energy.

The physical origin of the incommensurate magnetic fluctuation in
the doped bilayer triangular antiferromagnet is the almost same as
in the single layer case \cite{n14}, and is induced by the spinon
self-energy renormalization due to holons, {\it i.e.}, the
mechanism of the incommensurate type of structure away from the
half-filling is most likely related to the holon motion.
$S({\bf k},\omega)$ in Eq. (18) exhibits peaks when the incoming
neutron energy $\omega$ is equal to the renormalized spin
excitation $E_{k}^{2}=(\omega_{k}^{(1)})^{2}+B_{k}^{(1)}{\rm Re}
\Sigma_{LT}^{(s)}(k,E_{k})$, {\it i.e.},
$[\omega^{2}-(\omega_{k_{c}}^{(1)})^{2}-B_{k_{c}}^{(1)}{\rm Re}
\Sigma_{LT}^{(s)}(k_{c},\omega)]^{2}=(\omega^{2}-E_{k_{c}}^{2})^{2}
\sim 0$ for certain critical wave vectors ${\bf k}_{c}$. The height
of these peaks then is determined by the imaginary part of the
spinon self-energy $1/{\rm Im}\Sigma_{LT}^{(s)}(k_{c},\omega)$.
This renormalized spin excitation is doping and temperature
dependent. Near half-filling, the spin excitations are centered
around the antiferromagnetic wave vector ${\bf Q}$, so the
commensurate antiferromagnetic peak appears there. Upon doping, the
holes disturb the antiferromagnetic background. Within the
fermion-spin framework, as a result of self-consistent motion of
holons and spinons, the incommensurate antiferromagnetism is
developed beyond certain critical doping, which means that the
low-energy spin excitations drift away from the antiferromagnetic
wave vector, where the physics is dominated by the spinon
self-energy ${\rm Re}\Sigma_{L}^{(s)}(k,\omega)$ renormalization
due to holons. This is why the mobile holes are the key factor
leading to the incommensurate antiferromagnetism, while the spinon
energy dependence is ascribed purely to self-energy effects which
arise from the holon-spinon interaction. Since the height of the
incommensurate peaks is determined by damping, it is fully
understandable that they are suppressed as the neutron energy
$\omega$ and temperature are increased.

\section{Summary and Discussions}

In the above discussions, the central concern of the charge
transport and spin response in the doped bilayer triangular
antiferromagnet is the quasi-two dimensionality of the electron
state, then the charge transport is mainly determined by the
longitudinal charged holon fluctuation, and spin response is
dominated by the longitudinal spinon fluctuation. On the other
hand, our present study also indicates that the physical properties
of the doped antiferromagnet are heavily dependent on the electron
geometry structure. Unlike the doped square antiferromagnet, the
resistivity in the doped triangular antiferromagnet does not
exhibits a linear metallic-like behavior in the higher temperature
range as would be expected for the doped square antiferromagnet,
while six incommensurate neutron scattering peaks in the doped
triangular antiferromagnet are incompatible with these in the
doped square antiferromagnet, where four incommensurate peaks are
observed \cite{n5,n6}.

In summary, we have discussed the charge transport and spin
response of the doped bilayer triangular antiferromagnet within
the framework of the fermion-spin theory based on the $t$-$J$
model. It is shown that although the bilayer interaction leads to
the band splitting in the electronic structure, the qualitative
behaviors of the charge transport and spin response are the same as
in the single layer case \cite{n13,n14}. The conductivity spectrum
shows the low-energy peak and unusual midinfrared band separated by
the charge-transfer gap, while the temperature dependent
resistivity is characterized by the nonlinearity metallic-like
behavior in the higher temperature range, and the deviation from
the metallic-like behavior in the lower temperature range. The
commensurate neutron scattering peak near the half-filling is split
into six incommensurate peaks in the underdoped regime, with the
incommensurability increases with the hole concentration at lower
dopings, and saturates at higher dopings.

\acknowledgments

The authors would like to thank Feng Yuan and Jihong Qin for
helpful discussions. This work was supported by the National
Natural Science Foundation of China, and the special funds from
the Ministry of Science and Technology of China.

\end{document}